\documentclass[12pt,twoside]{article} 

\usepackage{bu_ece_report}
\usepackage{graphicx}
\usepackage{balance}  
\usepackage{balance}  
\usepackage{color}
\usepackage{authblk}
\usepackage{url}
\usepackage{supertabular,multicol}
\usepackage{colortbl}
\usepackage{changepage}
\usepackage{subfig}
\usepackage{float}
\definecolor{mygray}{gray}{.88}

\begin{document}

\buecedefinitions%
        {Identifying Dwarfs Workloads in Big Data Analytics}
        {}
        {Wanling Gao, Chunjie Luo, Jianfeng Zhan, Hainan Ye, Xiwen He, \\Lei Wang, Yuqing Zhu and Xinhui Tian}
        {Beijing, China}
        {May.\ 26, 2015}
        {2015-5} 

\buecereporttitlepage



\buecereportsummary{Big data benchmarking is particularly important and provides applicable yardsticks for evaluating booming big data systems. However, wide coverage and great complexity of big data computing impose big challenges on big data benchmarking. How can we construct a benchmark suite using a minimum set of units of computation to represent diversity of big data analytics workloads?
Big data dwarfs are abstractions of extracting frequently appearing operations in big data computing.
One dwarf represents one unit of computation, and big data workloads are decomposed into one or more dwarfs. Furthermore, dwarfs workloads rather than vast real workloads are more cost-efficient and representative to evaluate big data systems.
In this paper, we extensively investigate six most important or emerging application domains i.e. search engine, social network, e-commerce, multimedia, bioinformatics and astronomy. After analyzing forty representative algorithms, we single out eight dwarfs workloads in big data analytics other than OLAP, which are linear algebra, sampling, logic operations, transform operations, set operations, graph operations, statistic operations and sort.
}




\section{Introduction}  
The prosperity of big data and corresponding systems make benchmarking more important and challenging. Many researchers from academia and industry attempt to explore the way to define a successful big data benchmark. However, the properties of complexity, diversity and rapid evolution make us wonder where to start or how to achieve a wide range of coverage of diverse workloads. One attempt is to benchmark using popular workloads, which is very  subjective~\cite{ferdman2011clearing}; Another attempt is to focus on specific domains or systems~\cite{armstrong2013linkbench,ycsb,huang2010hibench}. These research efforts do not extensively analyze representativeness of workloads and fail to cover the complexity, diversity and rapid evolution of big data comprehensively.
The concept of dwarfs, which first proposed by Phil Colella~\cite{colella2004defining}, is thought to be a highly abstraction of workload patterns. To cover diversity of big data analytics, the dwarfs abstraction is of great significance.
First, it is a highly abstraction of computation and  communication patterns of big data analytics~\cite{asanovic2006landscape}; Second, it is a minimum set of necessary functionality~\cite{minimumSet}, which has strong expressive power, with one dwarf representing one unit of computation; Third, it is a direction for evaluation and performance optimization, e.g. guidelines for architectural research~\cite{asanovic2006landscape}.

Much previous work~\cite{codd1970relational,chen2014tpc,colella2004defining,asanovic2006landscape,shah2010data} has illustrated the importance of abstracting dwarfs in corresponding domains. TPC-C~\cite{chen2014tpc} is a successful benchmark which builds based on units of computation in OLTP domain. 
HPCC~\cite{luszczek2006hpc} adopts an analogous method to design a benchmark for high performance computing.
The National Research Council \cite{council2013frontiers} proposed seven giants in massive data analysis, which focus on major computational tasks or problems. 
These seven giants proposed by NRC are macroscopical definition of problems from the perspective of mathematics, rather than units of computation that frequently appeared in these problems.
Therefore, it is necessary to build a big data benchmark on top of dwarfs workloads which represent different units of computation. 
However, wide coverage and great complexity of big data impose great challenges to dwarfs abstraction.
1) There are massive application domains gradually involving big data. At present, big data has already infiltrated into all walks of life. Many domains have the requirements of storing and processing big data, and the most intuitive expression is billions of WebPages, massive remote sensing data, a sea of biological data, videos on YouTube, huge traffic flow data, etc.
2) In multiple research fields, there are powerful methods for big data processing. For great treasure hidden in big data, industrial and academic communities are both committed to explore effective processing methods, and now, many technologies have been successfully applied in above application domains, such as data mining, machine learning, deep learning, natural language processing, etc.
3) Large numbers of algorithms and the variants of these algorithms aggravate the difficulty of abstraction.
4) Not like traditional database systems, majority of big data are unstructured and operations on data are complicated.
To the best of our knowledge, none of existing big data benchmarks has identified dwarfs workloads in big data analytics.

In this paper, we propose the methodology of identifying dwarfs workloads in big data analytics, through a broad spectrum of investigation and a large number of statistical analysis.
We adopt an innovative and comprehensive methodology to investigate multi-field and multi-disciplinary of big data. At the first step, we singled out important and emerging application domains, using some widely acceptable metrics.
In view of the selected application domains, we investigated the widely used technologies in these domains  (i.e., machine learning, data mining, deep learning, computer vision, natural language processing, information retrieval) and existing libraries (i.e., Mahout \cite{mahout}, MLlib \cite{mllib}, Weka \cite{hall2009weka}, AstroML \cite{codd1970relational}), frameworks (i.e., Spark \cite{spark}, Hadoop \cite{hadoopweb}, GraphLab \cite{low2014graphlab}), benchmarks (i.e., BigBench \cite{bigbench}, AMP Benchmark \cite{amplabbenchmark}, LinkBench \cite{armstrong2013linkbench}, CALDA \cite{pavlo2009comparison}), which reflect the concerns of big data analytics.
Then at the third step, we singled out 40 representative algorithms. After analyzing these algorithms and summarizing frequently appearing operations, we finalized eight \emph{kinds} of workloads as the dwarfs workloads  in big data analytics.
In order to verify their accuracy and comprehensiveness, we analyzed typical workloads and data sets in each domain from two perspectives: diverse data models of different types (i.e., structured, semi-structured, and unstructured), and different semantics (e.g., text, graph, table, multimedia data);  We confirm through using a Directed Acyclic Graph(DAG)-like structure description, with an edge and a vertex to represent the dwarfs and the data set(or subset) respectively, we compose the original forty algorithms using combinations of one or more dwarfs workloads. 


Guided by the eight dwarfs workloads in big data analytics, we present an open-source big data benchmark suite called BigDataBench 3.1, with several industrial partners, publicly available at \url{http://prof.ict.ac.cn/BigDataBench/}. It is a significantly upgraded version of our previous work -- BigDataBench 2.0~\cite{wang2014bigdatabench}.
As a multi-discipline research and engineering effort spanning system, architecture, and data management, involving both industry and academia, the current version of BigDataBench includes 14 real-world data sets, and 33 big data workloads.


The rest of the paper is organized as follows. In Section 2, we describe the background, related work and motivation. Section 3 presents the methodology of abstracting dwarfs workloads in big data analytics and properties of these dwarfs. Section 4 states how dwarfs guide the construction of BigDataBench. Section 5 discusses the differences between our eight dwarfs and related work. Finally, we draw a conclusion in section 6.

\section{Background, Related Work and Motivation}

In 1970, E. F. CODD~\cite{codd1970relational} proposed a relational model of data, setting off a wave of relational database research, which is the basis of relational algebra and theoretical foundation of database, especially corresponding query languages. The \emph{set} concept in relational algebra abstracts five primitive and fundamental operators (Select, Project, Product, Union, Difference), which have fine expression, for different combinations can build different expression trees of queries.
Analogously, Phil Colella~\cite{colella2004defining} identified seven dwarfs of numerical methods which he thought would be important for the next decade. Based on that, a multidisciplinary group of Berkeley researchers propose 13 dwarfs which are highly abstraction of parallel computing, capturing the computation and communication patterns of a great mass of applications~\cite{asanovic2006landscape},  through identifying the effectiveness of the former seven dwarfs in other collections of benchmark, i.e. EEMBC, and three increasingly important application domains, i.e. machine learning, database software, and computer graphics and games.
There are still some successful benchmarks constructed based on abstraction.
TPC-C~\cite{tpcc} proposed the concepts of functions of abstraction and functional workload model, articulated around five kinds of transactions that frequently appeared in OLTP domain~\cite{chen2014tpc}, making it to be a popular yardstick.
HPCC~\cite{luszczek2006hpc} is a benchmark suite for high performance computing, which consists of seven basically tests, concentrating on different computation, communication and memory access patterns.
These successful stories demonstrate the necessity of constructing big data benchmarks based on dwarfs. 
With the booming of big data systems, diverse workloads with rapid evolution appear, making big data benchmarking difficult to achieve a wide coverage for a tough problem of workloads selection. 
In this condition, identifying the dwarfs workloads of big data analytics and building benchmarks based on these core operators become particularly important, moreover, optimizing these dwarfs workloads will have great impacts on performance optimization. 
This paper focuses on a fundamental issue---what are dwarfs workloads in big data analytics and how to find them? 

The National Research Council \cite{council2013frontiers} proposed seven major computational tasks in massive data analysis, which are called giants. There are great differences between those seven giants with our eight dwarfs.
1) They have different level of abstraction. NRC concentrates on finding major problems in big data analytics. In contrast, we are committed to decompose major algorithms in representative application domains and find units of computation that frequently appearing in these algorithms, which is at a lower level and more fine-grained.
2) Since they have different focuses, the results are also different. Most of the seven giants are a class of big problems. For example, generalized N-body problems are a series of tasks involving similarities between pairs of points, alignment problems refer to matchings between two or more data sets. However, our eight dwarfs are results of decomposition of main algorithms in big data analytics, which are summarized units of computation.
3) Combination of our eight dwarfs can compose algorithms which belong to above seven giants. That is to say, combination of dwarfs can be a solution for seven major problems. For instance, k-means involving similarities between points, which belongs to generalized N-body problem, is composed of vector calculations and sort operations.
Above all, the seven giants are macroscopical definition of problems from the perspective of mathematics, while our eight dwarfs are fine-grained decomposition of major algorithms in application domains and statistical analysis of these algorithms.
The differences will be further described in Section \ref{discussion}. 
In addition, Shah et al. \cite{shah2010data} discussed a data-centric workload taxonomy with the dimensions of response time, access pattern, working set, data types, and processing complexity, and proposed an example of key data processing kernels.

Big data attracts great attention, appealing many research efforts on big data benchmarking.
BigBench \cite{bigbench} is a general big data benchmark based on TPC-DS \cite{tpcds} paying attention to big data analytics and covering three kinds of data types.
HiBench \cite{huang2010hibench} is a Hadoop benchmark suite, which contains 10 Hadoop workloads, including micro benchmarks, HDFS benchmarks, web search benchmarks, machine learning benchmarks, and data analytics benchmarks.
YCSB \cite{ycsb} released by Yahoo! is a benchmark for data storage systems and only includes online service workloads, i.e. Cloud OLTP.
CALDA \cite{pavlo2009comparison} is a benchmarking effort for big data analytics.
LinkBench \cite{armstrong2013linkbench} is a synthetic benchmark based on social graph data from Facebook.
AMP benchmark \cite{amplabbenchmark} is a big data benchmark proposed by AMPLab of UC BerKeley which focus on real-time analytic applications.
Zhu et al. \cite{zhu2014bigop} proposed an benchmarking framework -- BigOP, abstracting data operations and workload patterns.

\section{Methodology}
\begin{figure}[!t]
\centering
\includegraphics[scale=0.6, angle=90]{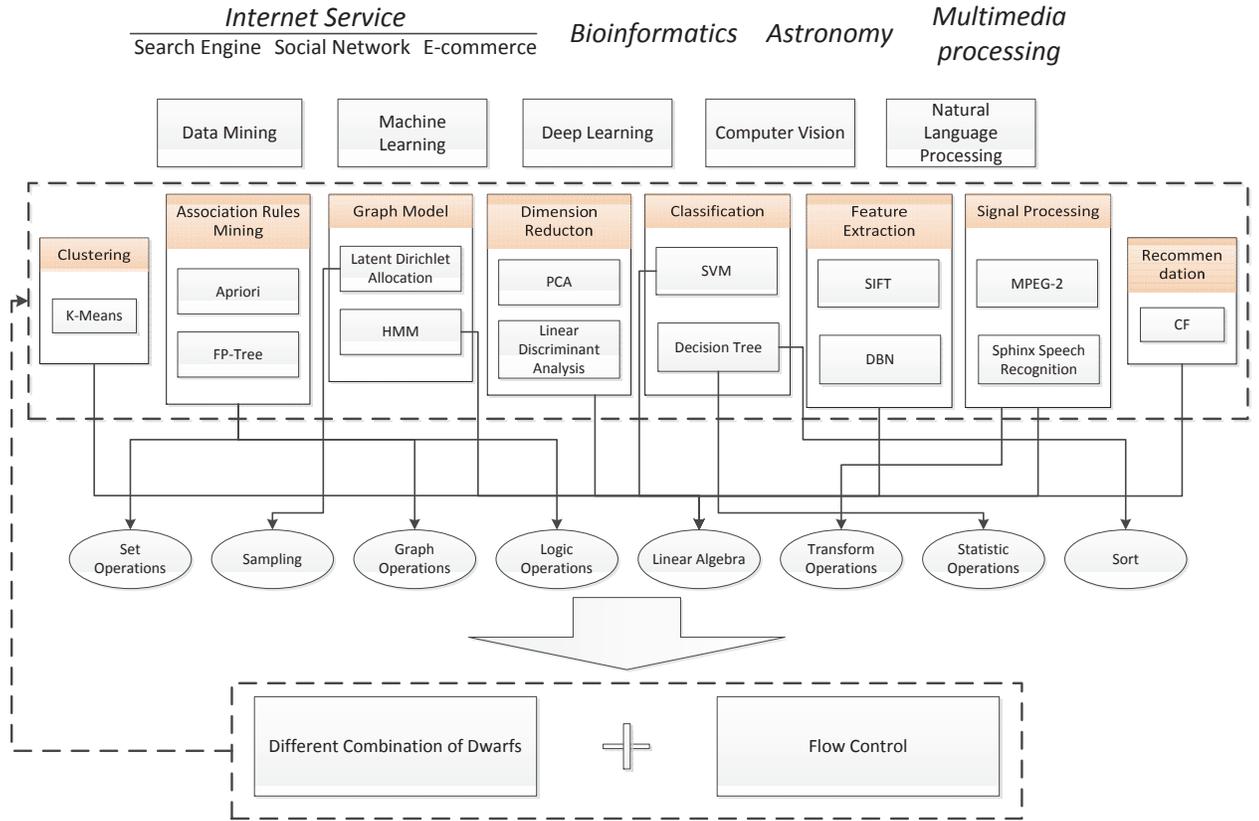}
\caption{Overall Structure of Dwarfs Abstraction.} 
\label{mapdwarfs}
\end{figure}

This section presents our methodology on dwarfs abstraction of big data analytics.
Before diving into the details of dwarfs abstraction methodology, we first introduce the overall structure.
Fig. \ref{mapdwarfs} illustrates the whole process of dwarfs abstraction and explains how algorithms map down to dwarfs.
We first investigate the main application domains and explore the widely used techniques, and then representative algorithms are chosen to summarize the frequently appearing operations, and finally conclude eight dwarfs using a statistical method.  We confirm a combination of  one or more dwarfs can compose the  40 original algorithms with different flow controls, e.g., iteration, selection.  

In the dwarfs abstraction of big data analytics, we omit the flow control of algorithm, i.e., iteration, and basic mathematical functions, i.e., derivative. The reason why we take these considerations is that our goal is to explore the dwarfs which appear frequently in algorithms, then we care more about the essence of computation instead of flow control.

\subsection{Dwarfs Abstraction Methodology}
\begin{figure}[!t]
\centering
\includegraphics[scale=0.6,angle=90]{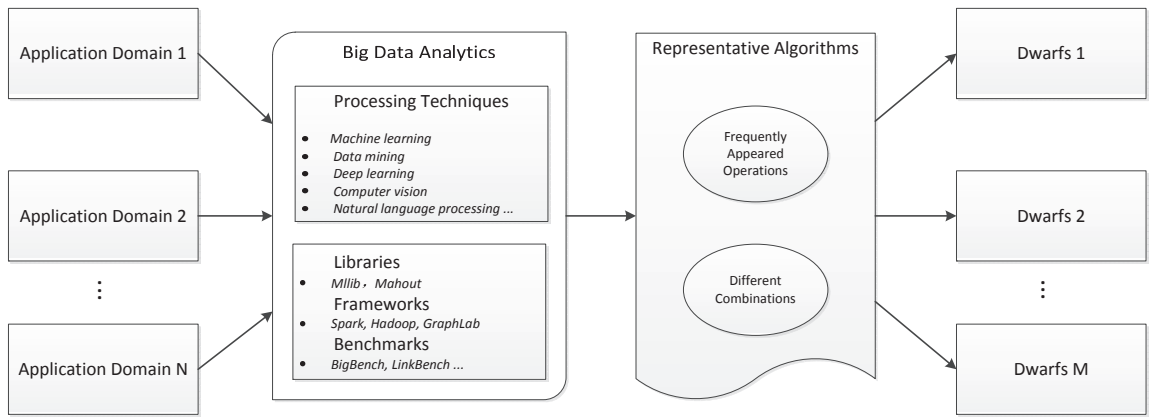}
\caption{Dwarfs Abstraction Methodology.} 
\label{DwarfsM}
\end{figure}

Dwarfs are highly abstractions of frequently appearing operations, and we adopt an innovative and comprehensive approach to abstract dwarfs of big data analytics, covering data models of different types (i.e., structured, semi-structured, and unstructured) and semantics (i.e., text, graph, table, multimedia data). Fig. \ref{DwarfsM} describes the methodology we use to abstract a full spectrum of dwarfs that are widely used in big data analytics.

Seltzer et al. \cite{seltzer1999case} pointed that we need use application-specific benchmarks to produce meaningful performance numbers in the context of real applications, and Chen et al. \cite{chen2014tpc} argued that the benchmark should measure performance using metrics which reflect real life computational demands and are relevant to real life application domains.
At the first step, we single out important and emerging application domains, using widely acceptable metrics.
To investigate the typical application domains of internet service, we use metrics of the number of page views and daily visitors , and further found out that 80\% page views of internet service came from search engine, social network and e-commerce \cite{alexatop500}. In addition, for the emerging and burgeoning domains, multimedia, bioinformatics and astronomy are three domains which occupied main positions in big data analytics \cite{multimediagrowth,genogrowth,astronomy}.


In allusion to selected application domains, we have the following two considerations. On one hand, big data analytics involves many advanced processing techniques; On the other hand, many open source tools for processing big data exist, such as libraries (i.e., MLlib \cite{mllib}, Mahout \cite{mahout}), frameworks (i.e., Spark \cite{spark}, Hadoop \cite{hadoopweb}, GraphLab \cite{low2014graphlab}), and a series of benchmark suites in some way reflect the concerns of big data analytics, such as BigBench \cite{bigbench}, LinkBench \cite{armstrong2013linkbench}. In view of the above two points, we choose representative algorithms widely used in data processing techniques, considering in conjunction with open source projects of libraries, frameworks and benchmarks.
After choosing representative algorithms which play important roles in big data analytics, we deeply analyze the process and dig out frequently appearing operations in these algorithms. Moreover, different combinations of operations are considered to compose original algorithms.
Finally, we summarize the dwarfs workloads in big data analytics.
A Directed Acyclic Graph (DAG)-like structure is used to specify how data sets (or subsets) are operated by dwarfs.


%

\subsection{Algorithms Chosen to Investigate}
Data is not the same thing as knowledge, however, data can be converted into knowledge after being processed and analyzed, which needs powerful tools to digest information.
We analyze the process of the above-mentioned application domains with the purpose of singling out representative algorithms in these six domains. There are generality and individuality among  different domains.

\begin{figure}[!t]
\centering
\includegraphics[height=4.2in, width=7in]{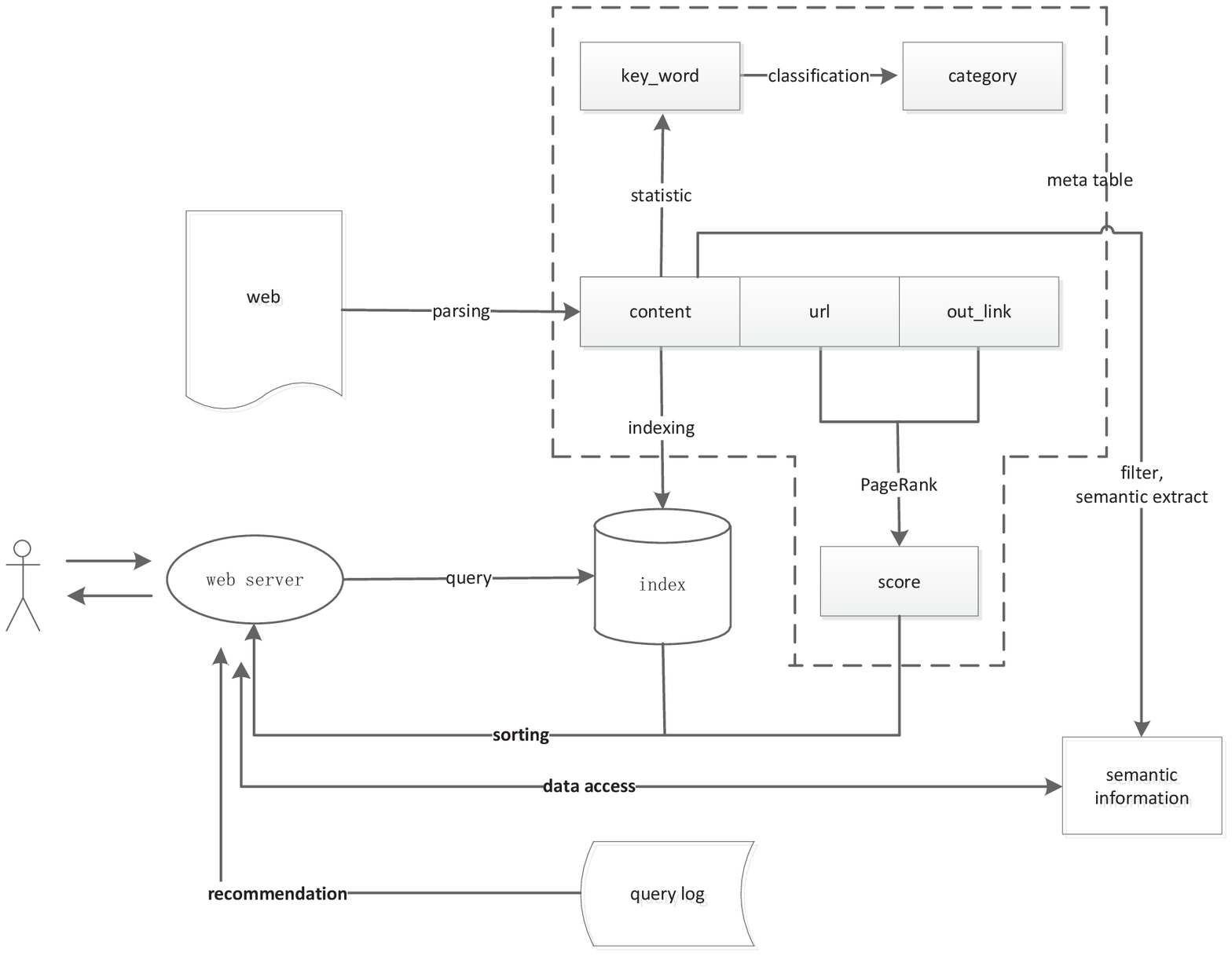}
\caption{Process of a search engine.}
\label{search_engine}
\end{figure}

Taking search engine as an example, we illustrate how we choose algorithms according to a selected application domain.
Fig. \ref{search_engine} shows the details of search engine. After obtaining the web pages from spider, the parser extracts the text content and clears the structure of the web graph. Then
several analysis methods are executed, including not only analysis on text content (statistic, index, semantic extract, classification), but also on web graph (pagerank). Moreover, query recommendation \cite{he2009web,zhang2006mining,li2008pfp} is provided in case of unfamiliarity with terminology or dissatisfaction with results.
After analyzing several necessary algorithms which construct search engine, we choose the following algorithms for investigation, including index, porter stemming, pagerank, HITS, classification (decision tree, naive bayes, svm, etc), recommendation and semantic extract (latent semantic indexing, latent dirichlet allocation), covering many technologies, such as data mining, machine learning.

In fact, most algorithms are not only used in one application domain, but also applied to other domains. Taking aforementioned classification methods as an example, they have been widely used in the other five domains under investigation.
After conducting an thoroughly survey based on the six application domains, we also refer to the  top 10 algorithms \cite{wu2008top} and 18 candidates \cite{18candidate} in data mining, and several machine learning algorithms covering classification, regression, clustering, dimension reduction, recommendation, and computer vision algorithms spanning from a function i.e. feature exaction to applications components (i.e., image segmentation, ray tracing). The other algorithms include  classic deep learning algorithms and sequence alignment algorithms, both of which have  a broad range of application. We also include  important algorithms in mainstream libraries (i.e., OpenCV, MLlib, Weka, AstroML), frameworks (i.e., Spark, Hadoop, GraphLab) and implemented workloads in benchmarks (i.e., BigBench, LinkBench, AMP Benchmark, CALDA). In total, we choose 40 widely used algorithms to investigate. The algorithms are listed in Table. \ref{algorithms} from perspectives of typical application domain, brief description, frequently-appearing operations. 


\begin{table}[htbp]
\caption{Investigated Algorithms.}\label{algorithms}
\center
 \begin{tabular}{|p{1.3in}|p{2.0in}|p{3.0in}|}
 \hline \rowcolor{mygray} \textbf{Investigated Algorithms}  &   \textbf{Brief Description}  & \textbf{Frequently-appearing Operations}    \\ \hline
 \multicolumn{3}{|l|}{}\\
 \multicolumn{3}{|l|}{\textbf{Data Mining \& Machine Learning}}\\
 \hline
  C4.5/CART/ID3   & Decision tree for classification or regression & Count numbers to computer information gain or Gini coefficient; Sort for splitting attribute; Build or prune tree\\
  \hline
  Logistic Regression & A method for classification or regression using logistic function, with the output between 0 and 1  &Vectorization of gradient descent method; Matrix operations(multiplication, transposition, inverse) using formula\\
  \hline
  Support Vector Machine (SVM) & A supervised learning method for classification or regression, maximal margin classifier & Vector multiplication; Kernel function\\
  \hline
  k-Nearest Neighbors Algorithm (k-NN) & A non-parametric method for classification or regression & Similarity calculation of vectors; Sort to find k nearest neighbors; Count the number of categories\\
  \hline
  Naive Bayes & A probabilistic classifier based on Bayes' theorem &Count for probability calculation\\
  \hline
  Hidden Markov Model (HMM)  & Generating a model assuming the hidden variables to be a Markov process &Matrix/Vector operations; Transfer-probability matrix\\
  \hline
  Maximum-entropy Markov Model (MEMM) & A discriminative graphical model used for sequence labeling &Matrix/Vector operations; Conditional transfer-probability matrix\\
  \hline
  Conditional Random Field (CRF)  & A probabilistic graphical model used for sequence labeling &Matrix/Vector operations; Compute normalized probability in the global scope \\
  \hline
  PageRank& An algorithm used for ranking webpages &Matrix operations (multiplication, transpose)\\
  \hline
  HITS &An algorithm used for ranking webpages based on Hubs and Authorities&Authority and hubness vector of webpages; Link matrix; Matrix-vector multiplication\\
  \hline
  Aporiori  &Mining frequent item sets and learning association rules &Set operations(intersection); Count the number of items; Hash tree\\
  \hline
  FP-Growth & Mining frequent item sets using frequent pattern tree &Set operations(intersection); Count the number of items; Build tree; Sort according to support threshold\\
  \hline
\end{tabular}
\end{table}

\begin{table}[htdp]
\center
 \begin{tabular}{|p{1.3in}|p{2.0in}|p{3.0in}|}
 \hline
  K-Means  & A clustering method determined by the distances with the centroid of each cluster & Similarity calculation of vectors; Sort \\
 \hline
 Principal Component Analysis (PCA) &A unsupervised learning method used for dimensionality reduction &Solve the covariance matrix (matrix multiplication and transposition) and corresponding eigenvalue and eigenvector; Sort eigenvector according to eigenvalue\\
  \hline
  Linear Discriminant Analysis&A supervised learning method for classification&Covariance matrix; Vector operations (Transpose, subtraction, multiplication); Solve eigenvalue and eigenvector; Sort the maximum eigenvector according to eigenvalue\\
  \hline
  Back Propagation & A supervised learning method for neural network & Matrix/Vector operations (multiplication); Derivation\\
  \hline
  Adaboost  & A strong classier composed of multiple weak weighted classifiers &Train to get weak classifier (i.e., decision tree); Count the number of misclassified train data; Recompute weight distribution of train data\\
  \hline
  Markov Chain Monte Carlo (MCMC) & A series of algorithms for sampling from random distribution & Sampling\\
  \hline
  Connected Component (CC) & Computing connected component of a graph &BFS/DFS; Transpose graph; Sort the finishing time of vertexes\\
  \hline
  Random Forest & A classifier consists of multiple decision trees & Random sampling; Decision Tree \\
  \hline
   \multicolumn{3}{|l|}{}\\
  \multicolumn{3}{|l|}{\textbf{Natural Language Processing}}\\
 \hline
  Latent Semantic Indexing (LSI)  & An indexing method to find the relationship of words in huge amounts of documents & SVD; Count for probability calculation\\
  \hline
  pLSI  & An method to analyze co-occurrence data based on probability distribution & EM algorithm; Count to compute probability\\
  \hline
  Latent Dirichlet Allocation & A topic model for generating the probability distribution of topics of each document &Gibbs sampling/ EM algorithm; Count to compute probability\\
  \hline
   Index & Building inverted index of documents to optimize the querying performance &Hash; Count for probability calculation; Operations in HMM, CRF for Segmentation; Sort\\
  \hline
  Porter Stemming  &Remove the affix of words to get root &Identify the consonant and vowel form of words; Count the number of consonant sequences; stem suffix according to rules\\
  \hline
  \end{tabular}
\end{table}

\begin{table}[htdp]
\center
 \begin{tabular}{|p{1.3in}|p{2.0in}|p{3.0in}|}
 \hline
  Sphinx Speech Recognition & Translating the input audio into text &Operations in HMM; FFT; Mel-frequency cepstral coefficient; Vector representation of audio signal\\
  \hline
  \multicolumn{3}{|l|}{}\\
 \multicolumn{3}{|l|}{\textbf{Deep Learning}}\\
  \hline
  Convolution Neural Network (CNN)  & A variation of multi-layer perceptrons &Convolution; Subsampling; Back propagation\\
  \hline
  Deep Belief Network (DBN)  & A generative graphical model consists of multiple layers&Contrastive divergence; Gibbs sampling; Matrix/Vector operations \\
  \hline
  \multicolumn{3}{|l|}{}\\
   \multicolumn{3}{|l|}{\textbf{Recommendation}}\\
  \hline
  Demographic-based Recommendation & Recommending might interested items to one user based on their similarity to other users &Similarity analysis of user model\\
  \hline
  Content-based Recommendation  &Recommending might interested items to one user based on these items' similarity to previous bought items of the user &Similarity analysis of item model\\
  \hline
  Collaborative Filtering (CF) & Predicting the items which might be interested by specific users &Similarity calculation of vectors; QR decomposition\\
  \hline
\multicolumn{3}{|l|}{}\\
  \multicolumn{3}{|l|}{\textbf{Computer Vision}}\\
  \hline
  MPEG-2  &International standards of video and audio compression proposed in 1994 &Discrete cosine transform; Sum of Absolute Differences(matrix subtraction); Quantization matrix; Variable length coding(sort the frequency of the input sequence, binary tree)\\
 \hline
 Scale-invariant Feature Transform (SIFT) & An algorithm to detect and describe local features in images &Convolution; Downsampling; Matrix subtraction; Similarity calculation of vectors; Sort; Count\\
  \hline
  Image Segmentation (GrabCut) & Partitioning an image into multiple segments & Gaussian Mixture Model; Matrix operations(covariance matrix, inverse matrix, determinant, multiplication); Similarity calculation of pixels; K-means; Graph algorithms(MaxFlow, Min-cut) \\
  \hline
  Ray Tracing  & A rendering method for generating an image through tracing the path of light & Set operations(intersection); Hash; Vector representation of points\\
  \hline
\end{tabular}
\end{table}

\begin{table}[htdp]
\center
 \begin{tabular}{|p{1.3in}|p{2.0in}|p{3.0in}|}
 \hline
  \multicolumn{3}{|l|}{}\\
 \multicolumn{3}{|l|}{\textbf{Database Software}}\\
  \hline
  Needleman-Wunsch  & An dynamic programing algorithm for global sequence alignment&Count the length of sequences; Computing scoring matrix; Backtrace from the bottom right corner of the matrix\\
  \hline
  Smith-Waterman &An dynamic programing algorithm for local sequence alignment&Count the length of sequences; Computing scoring matrix; Sort for the largest score value in the matrix; Backtrace from the largest value until the score is zero\\
  \hline
  BLAST  & An heuristic approach for sequence alignment &Score matrix; Sort for pairs of aligned residues higher than threshold; Hash table; Seeding-and-extending \\
  \hline
\end{tabular}
\end{table}

After investigating the 40 algorithms, we analyzed their frequently appearing operations and identified eight dwarfs workloads.
As summarized in Table. \ref{algorithms}, linear algebra plays a fundamental  role in algorithms effectively for big data analytics, for many problems can be abstracted into matrixes or vectors operations, such as SVM, K-means, PCA, CNN, etc. In addition, most graph-theoretical problems can be converted to matrix computations, for HMM (Probabilistic graphical model), PageRank (Webgraph), etc.
Other graph-theoretical problems include graph traverse problem, such as BFS, shortest path problems, etc.
Many investigated algorithms involve in similarity measurement, i.e. k-NN, collaborative filtering. Common similarity calculation methods include Euclidean distance, Manhattan distance, Jaccard similarity coefficient, etc. Most of these methods focus on basic vector calculation, while jaccard similarity coefficient adopts the concept of set, using the number of the intersection divided by the number of the union of the input sets, which is also applied to a large class of algorithms for association rules mining i.e., apriori, fp-growth and theory of rough set and fuzzy set. In addition, the main operations in relation algebra are set operations.

The PageRank algorithm which makes Google rise to fame,  applies one category of sampling (markov chain monte carlo) methods in  prediction the next page visited, which forms a markov chain. Not only that, sampling methods have an significant position in many algorithms and applications, i.e., boostrap, latent dirichlet allocation, simulation, boost, stochastic gradient descent.

The widespread use of transform operations in signal processing and multimedia processing greatly simplifies the computation complexities, for difficult computations in original domain can be easily computed in converted domain, such as FFT and DCT for MPEG, speech recognition. Furthermore, as seen in Table. \ref{algorithms},  convolution calculations play important roles, while FFT is an lower complexity implementation of convolution according to convolution theorem.
Another category of operations is hash, widely used in encrpytion algorithms, index, and fingerprint for similarity analysis.
There are still two primitive operations which are used in almost all the algorithms -- sort, statistics(i.e. count, probability calculation).

\subsubsection{Dwarfs Workloads}

In summary, Table. \ref{Kernels} lists dwarfs workloads widely used in big data analytics.

\begin{table}[!t]
\caption{Dwarfs in Big Data Analytics.}\label{Kernels}
\center
\begin{tabular}{|p{0.2in}|p{2.0in}|p{3.5in}|}

  \hline
  No.   &Operations            &   Description \\
  \hline
  1 & Linear Algebra    & Matrix/Vector operations, i.e., addition, subtraction, multiplication\\
  \hline
  2 &Sampling            & MCMC(i.e., Gibbs sampling), random sampling\\
  \hline
  3 & Logic Operations       &  A collection of Hash algorithms, i.e., MD5\\
  \hline
  4 & Transform Operations       &  FFT, DCT, Wavelet Transform\\
  \hline
  5 & Set Operations       &  Union, intersection, complement \\
  \hline
  6 & Graph Operations                   & Graph-theoretical computations, i.e., graph traversal\\
  \hline
  7 & Sort         & Partial sort, quick sort, top k sort \\
  \hline
  8 & Statistic Operations               & Count operations\\
  \hline
\end{tabular}
\end{table}

\textbf{Linear Algebra} In big data analytics, matrixes or vectors are without doubt a sharp weapon to solve many problems. From a dimension point of view, matrix operations consist of three categories, e.g. vector-vector, vector-matrix, matrix-matrix; From a storage standpoint, matrix operations are divided into two categories: sparse matrix and dense matrix.
The concrete operations of a matrix are primarily addition, subtraction, multiplication, inversion, transposition, etc.

%
%
%

\textbf{Sampling} Sampling is an essential step in big data processing.
Considering the following situation, if the exact solution of one problem can not be solved using analytical method, what other alternative do we have? To solve this problem, people attempted to get an approximate solution, approaching to the exact solution as far as possible. Stochastic simulation is an important category of methods in approximation analysis, and its core concept is sampling, including random sampling, importance sampling, markov chain monte carlo sampling, etc.

\textbf{Logic Operations} Hash is of great importance in a very wide range of computer applications, e.g., encryption, similarity detection and cache strategy in distributed applications. Hash can be divided into two main types including locality sensitive hash (LSH) and consistent hash. In multimedia area, LSH can be used to retrieve images and audio. Every image can be expressed by one or more feature vectors, through creating indexes for all the feature vectors, and the speed of similar image retrieval can be improved significantly. Moreover, it can be applied to duplicated web pages deletion and fingerprint matching, such as SimHash, I-Match, shinging, etc.

%

\textbf{Transform Operations} The transform operations here means the algorithms used in audio signal analysis, video signal processing and image transformation. Common algorithms are discrete fourier transform (DFT) and its fast version --- fast fourier transform (FFT), discrete cosine transform (DCT) and wavelet transform.

\textbf{Set Operations} In mathematics, set means a collection of distinct objects. Likewise, the concept of set can be applied to computer science. Set operations include union, intersection, complement of two data sets.
The most familiar application type which benefits from set operations is SQL-based interactive analysis. In addition, similarity analysis of two data sets involves set operations, such as Jaccard similarity. Furthermore, both fuzzy set and rough set play very important roles in computer science. Fuzzy set can be used to perform grey-level transformation and edge detection of an image.

\textbf{Graph Operations} A large class of applications involve graphs.
One representation of graph is matrix, then many graph computing problems convert to linear algebra computations.  Graph problems often involve graph traversing and graph models. 
Typical applications involving  graphs are social network, probabilistic graphic models, depth/breadth-first search, etc.

\textbf{Sort} Sorting is extensive in many areas. Jim Gray thought sort is the core of modern databases~\cite{asanovic2006landscape}, which shows its fundamentality. Even though in other domains, sort still plays a very important role.

\textbf{Statistic Operations} As with sort, statistic operations are also at the heart of many algorithms, such as probability or TF-IDF calculation.


\subsection{Properties of Dwarfs}
Dwarfs of big data analytics represent frequently appeared operations in algorithms for processing big data. They have some properties.\vspace{1ex}

 \begin{adjustwidth}{2ex}{}
\noindent \emph{\textbf{Composability}}: Algorithms for big data analytics are composed of one or several dwarfs, with certain flow control and basic mathematical functions. An DAG-like description are used to describe the process.\vspace{1ex}

 \noindent \emph{\textbf{Irreversibility}}:  The combination is sensitive to the order of dwarfs for a specific algorithm. Different combinations would have great impacts on performance or even produce different results.\vspace{1ex}

\noindent \emph{\textbf{Uniqueness}}: These eight dwarfs represent different computation and communication patterns in big data analytics.\vspace{1ex}
\end{adjustwidth}

\begin{figure}[!t]
\centering
\includegraphics[scale=0.7, angle=90]{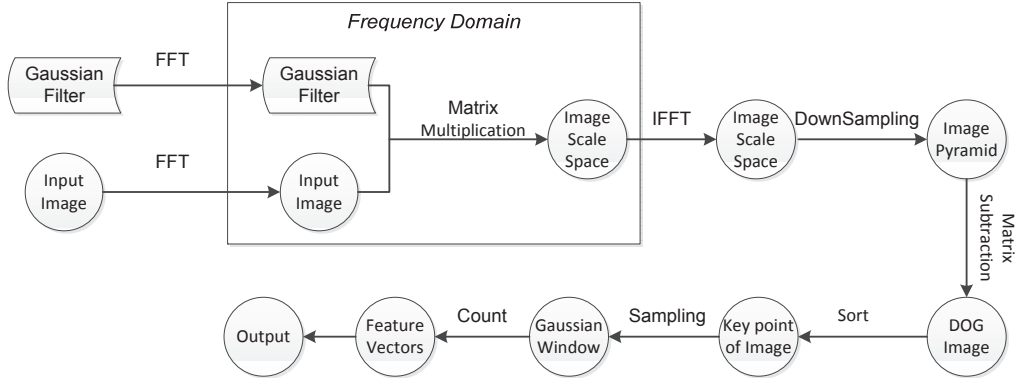}
\caption{The DAG-like Structure of SIFT Algorithm. SIFT as a representative algorithm in computer vision, is decomposed into several dwarfs workloads.} 
\label{SIFT}
\end{figure}

These dwarfs simplify the complexity of big data analytics, and they have strong expression power in terms that they can be combined into various algorithms.
We use a DAG-like structure, in which a node represents original data set or intermediate data set being processed, and an edge represents a kind of dwarfs. We have used DAG-like structure to understand existing benchmarks on big data analytics. Taking SIFT as an example, we explain why the eight dwarfs make sense. SIFT is an algorithm to detect and describe local features in input images which first proposed by D. G. Lowe in 1999  ~\cite{lowe2004distinctive}, involving several dwarfs. As illustrated in Fig. \ref{SIFT}, a DAG-like structure specifies how data set or intermediate data set are operated by different dwarfs.

An image can be represented as a matrix in the computer, with a matrix element representing one pixel point. Gaussian filter is an convolution kernel in accordance with gaussian distribution function, which is actually a matrix. Image scale space $L(x,y,\partial)$ is produced from the convolution of the gaussian filter  $G(x,y,\partial)$ with the input image$I(x,y)$, $\partial$ is space scale factor. According to convolution theorem, FFT is one fast implementation method for convolution, in this regard, we don't add convolution to our list of dwarfs though it is of great significance, especially in image processing. By setting different value of $\partial$, we can get a group of image scale spaces. Image pyramid is the consequence of downsampling these image scale spaces. DOG image means difference-of-Gaussian image, which is produced by matrix subtraction of adjacent image scales of each octave in image pyramid.
After that, every point in one DOG scale space would sort with eight adjacent points in the same scale space and points in adjacent two scale spaces, to find the key points in the image. Through computing the mold and direction of each key point and sampling in adjacent gaussian window, following by sort and statistic operations, we can get the feature vectors of the image.

\section{Big Data Benchmarking}
In this section, we describe how we apply the eight dwarfs to construct a big data benchmarking suite -- BigDataBench.

\begin{figure}[!t]
\centering
\includegraphics[scale=0.9, angle=90]{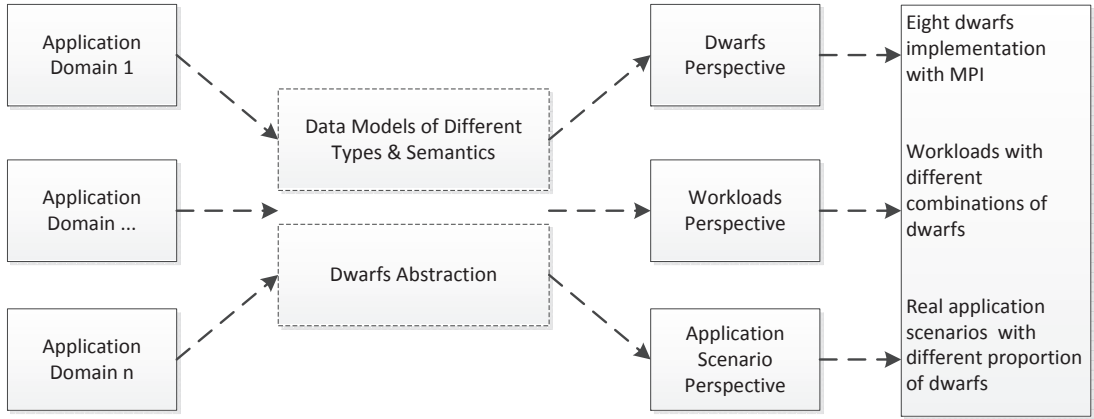}
\caption{Constructing BigDataBench Based on Dwarfs.} 
\label{bigdatabench}
\end{figure}

Fig. \ref{bigdatabench} shows the process of constructing BigDataBench using eight dwarfs.
We build big data benchmark from three perspectives: 1) Dwarfs perspective. Every dwarf is a collection of algorithms with similar patterns. For example, linear algebra contains many algorithms like matrix addition, multiplication, etc. We implement workloads of each dwarf with MPI, for it is much more lightweight then the other programming frameworks in terms of binary size.
 2) Workloads perspective. From the methodology of dwarfs abstraction, we single out representative workloads with different combinations of eight dwarfs, including 14 real-world data sets and 33 workloads. 
 3) Application scenario perspective. We also provide the whole application scenarios with different proportion of eight dwarfs. 
 
\section{Comparison with NRC Seven Giants}\label{discussion}

National Research Council proposed seven major tasks in massive data analysis~\cite{council2013frontiers}, which they called giants. These seven giants are basic statistics, generalized N-body problems, graph-theoretic computations, linear algebraic computations,  optimization, integration, and alignment problems.

\begin{figure}[!t]
\centering
\subfloat[NRC Seven Giants]{
\label{subfig_a}
\begin{minipage}[t]{0.25\textwidth}
\centering
\includegraphics[scale=0.9, angle=90]{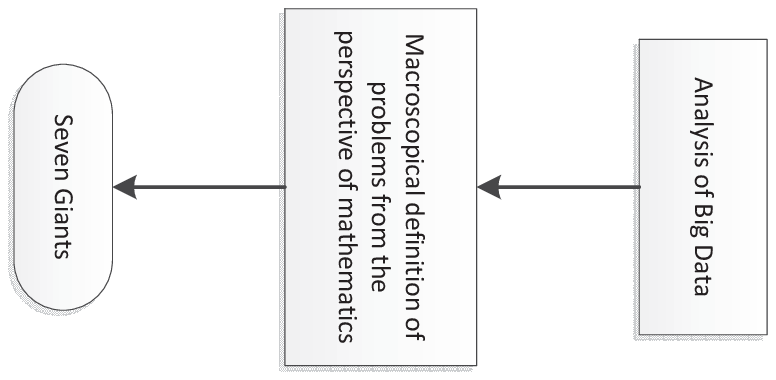}
\end{minipage}
}
\subfloat[Our Eight Dwarfs]{
\label{subfig_b}
\begin{minipage}[t]{0.25\textwidth}
\centering
\includegraphics[scale=0.8, angle=90]{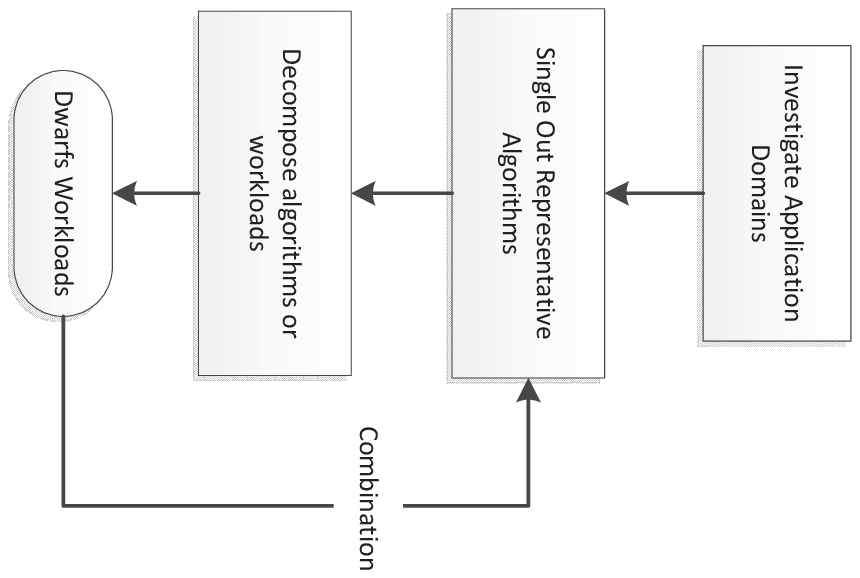}
\end{minipage}
}
\caption{Comparison of Identifying Methodology}
\label{comparison}
\end{figure}

In this section, we discuss the differences between our eight dwarfs and the NRC seven giants. Fig. \ref{comparison} lists our differences of identifying methodology. 
Fig. \ref{subfig_a} shows the process of summarizing seven giants. They focus on common used tasks and problems in massive data analysis, and then cluster them to identify seven giants. 
In this case, some giants are big problems, e.g. n-body problems, and some giants have a lot of overlap, for example, linear algebraic computations are a special case of optimization problems~\cite{council2013frontiers}. 
Fig. \ref{subfig_b} presents our methodology of identifying dwarfs workloads in big data analytics. We first choose representative application domains and corresponding processing techniques, then we analyze these advanced processing techniques and major open source projects to find representative algorithms in them. Next, we decompose these algorithms and summarize frequently appearing operations. At last, we finalize eight dwarfs workloads in big data analytics.
Our eight dwarfs are a lower level abstraction, which focus on units of computation in above tasks and problems. For example, a combination of one or more dwarfs with certain flow control can implement an optimization problem. 
Note that basic statistics, linear algebraic computations, and graph-theoretic computations are fundamental solutions for many problems, we also add them in our eight dwarfs.

\begin{adjustwidth}{2ex}{}
\noindent \emph{\textbf{Generalized N-body Problems}}: This category contains problems involving similarities between pairs of points, such as nearest-neighbor search problems, kernel summations. 
Our investigation partly covers algorithms in this category. For example, a class of algorithms for similarity analysis such as k-Nearest Neighbors algorithm and clustering methods such as k-means algorithm concern with similarity calculation of vectors (points), which is a large family of generalized N-body problems. Moreover, kernel summations such as support vector machine algorithm are also investigated.\vspace{1ex}

 \noindent \emph{\textbf{Optimization}}: This is a giant heavily relied on flow control. With several rounds of iteration, the result gradually converge to an extremum value.
 Optimization methods as a big class of mathematics, play an important role in computer science. In machine learning, the training models are learned through optimization procedures, such as neural network, support vector machine, adaboost, etc. In natural language processing, significant algorithms such as conditional random field adopt optimization methods to train parameters. 
 Our eight dwarfs omit the flow controls and concentrate on units of computation. However, they are important components of computational procedures in each iteration. 
For example, neural network algorithm is an optimization problem, but its each iteration is  linear algebraic computations.\vspace{1ex}

\noindent \emph{\textbf{Integration}}: It is a very important branch of mathematics. Integration are widely used in many problems, such as expectations and probability calculation. Markov chain monte carlo as one type of sampling, which is one of our eight dwarfs, has been applied to integration problems for an approximate solution according to the law of large numbers. \vspace{1ex}

\noindent \emph{\textbf{Alignment Problems}}: This class includes problems about matchings. Typical alignment problems are sequence alignment in bioinformatics, image features matching in multimedia area, which are also considered in our analysis, such as BLAST, scale-invariant feature transform.\vspace{1ex}
\end{adjustwidth}

\section{Conclusions}

In this paper, we identified eight dwarfs in big data analytics other than OLAP, through a broad spectrum of investigation and a large number of statistical analysis. We adopt an innovative methodology of singling out typical application domains (i.e., search engine, social network, e-commerce, bioinformatics, multimedia, and astronomy) at the first step. Then we focus on different algorithms widely used in these application domains and existing libraries, frameworks, benchmarks for big data analytics. After investigating these techniques and open source projects, we choose forty representative algorithms which play a significant role in big data analytics. Through deeply analyzing these algorithms and digging out the frequently appearing operations, we identify eight dwarfs workloads taking redundancy and comprehensiveness into consideration.



\parskip=0pt
\parsep=0pt
\bibliographystyle{ieeetrsrt}


\end{document}